\documentstyle [12pt] {article}
\begin{document}
\rightmargin -2.75cm
\textheight 23.0cm
\topmargin -0.5in
\baselineskip 16pt
\parskip 18pt
\parindent 30pt
\title{ \large \bf Fast Neutrino Decay in The Minimal
\\Seesaw Model}
\author{Anjan S. Joshipura \and Saurabh D. Rindani\\Theory Group, Physical
Research Laboratory\\
Navrangpura, Ahmedabad 380 009, India}
\date{}
\pagestyle{empty}
\pagenumbering{arabic}
\baselineskip 24pt
\maketitle
\vspace{.2in}
\baselineskip 12pt
\vspace{0.5cm}
\pagestyle{empty}
\baselineskip 16pt
\parskip 16pt
\pagestyle{plain}
\begin{abstract}
Neutrino decay in the minimal seesaw model containing three
right handed neutrinos and a complex $SU(2)\times U(1)$ singlet
Higgs  in addition to the standard model fields is considered. A
global horizontal symmetry $U(1)_H$ is imposed, which on
spontaneous breaking gives rise to a Goldstone boson. This symmetry
 is chosen in a way that makes a) the contribution of heavy ($\leq$
 MeV) majorana
neutrinos  to the neutrinoless double beta decay amplitude vanish
and b) allows the heavy neutrino to decay  to a lighter
neutrino and  the Goldstone boson. It is shown that this decay can
occur at a rate much faster than in the original Majoron model even
 if one does not introduce any additional Higgs fields as is done in
the literature. Possibility of describing the 17 keV neutrino in
this minimal seesaw model is investigated. While most of the
cosmological and astrophysical constraints on the 17 keV
neutrino can be satisfied in this model, the laboratory limits
coming from the neutrino oscillations cannot be easily met. An
extension which removes this inadequacy and offers a consistent
description of the 17 keV neutrino is discussed.
\end{abstract}
\newpage
\section{Introduction}
Apart from playing an important role in cosmology and
astrophysics, neutrino masses would provide an unambiguous
signature of physics beyond the minimal $SU(3)\times SU(2)\times
U(1)$ model. Various mechanisms for generating these masses in
gauge theories \cite{pal} have been proposed. Among them, the
seesaw mechanism of Gell-Mann, Ramond Slansky and Yanagida
(GRSY) \cite{GRSY} explains their smallness compared to other
fermions in a natural way and has been extensively studied. The
presence of (at least) one right-handed neutrino with a large
(Majorana) mass $\hat{M}$ is an essential ingredient in the GRSY
mechanism. This large mass $\hat{M}$ suppresses the masses of
ordinary left-handed neutrios which are typically given by
\begin{equation}
m_{\nu_i}\sim m^2_i/\hat{M},
\end{equation}
$i$=1,2,3 being the generation indices and $m_i$ is the Dirac
mass connecting the left- and the right-handed neutrinos of the
$i$th generation. $\hat{M}$ is related to the breaking of GUT or
left-right symmetry in many models, but, in general, it could
assume any value from $O(\rm{GeV})$ to $M_{Planck}$ in a
phenomenologically consistent manner. $m_i$ could be linked to
masses of the charge-$\frac{2}{3}$ quarks $u_i$ or to those of
the charged leptons $e_i$. One interesting aspect of the GRSY
mechanism is that there exists a range of natural values of
parameters $m_i$ and $\hat{M}$ for which the neutrino masses are
in the observable range and near their experimental limits.  For
example, if $\hat{M}$ is identified \cite{luty}with the weak
scale ($\approx$ 100 GeV) and $m_i$ with the charged lepton
masses, then $m_{\nu _e}\approx$ 2.5 eV, $m_{\nu _\mu}\approx$
100 keV, and $m_{\nu _\tau}\approx$ 30 MeV. Likewise, if $m_i$
are identified with the masses of $u_i$, then one could have
observable mass at least for $\nu _\tau$, even for a large
$\hat{M}$, e.g., $\hat{M}\approx 10^7$ GeV and $m_t\approx $ 100
GeV implies $m_{\nu_\tau}\approx $1 MeV.

Seesaw models which predict $m_{\nu _{\mu}}$ and $m_{\nu _\tau}$
in the above range have to satisfy two important constraints.
The first comes from neutrinoless double beta decay
(0$\nu\beta\beta$) which requires the effective mass $m_{eff}$
for $\nu _e$ to be less than a few electron volts. A heavy
neutrino mixing with $\nu _e$ could contribute significantly to
this effective mass, e.g., a neutrino with 1 MeV mass
contributes $>$O(eV) to $m_{eff}$ if its mixing $\beta$ with
$\nu _e$ is $> 10^{-3}$. One must avoid large contribution to
$0\nu
\beta\beta$ coming in this way from heavy masses. The second
constraint comes from the neutrino contribution to the energy
density of the universe \cite{turn}. In terms of masses, this
requires the sum of the masses of the normal neutrinos to be
$\leq$100 eV. Hence $\nu _\mu$ and $\nu _\tau$ have either to
decay or to annihilate in the early  universe if their masses
are large compared to this limit.  Both decay as well as
annihilation could occur if the theory  contains a massless
Goldstone boson, viz., the Majoron. Such a Goldstone boson can
be naturally incorporated \cite{cmp} in the seesaw model by
introducing a complex $SU(2)\times U(1)$ singlet Higgs field
whose vacuum expectation value sets the scale $\hat{M}$ and
breaks lepton number spontaneously, generating the Majoron. A
minimal model of this type is however shown \cite{sv} to be
inadequate for generating fast decay of the heavy neutrino. A
typical decay rate for the process $\nu _{h}\rightarrow \nu
_l+J$ is shown \cite{sv} to be
\begin{equation}
\Gamma \approx \frac{\beta ^2}{16\pi}\frac{m^5_{\nu_h}}{\hat{M}^4}.
\end{equation}
The life time implied by the above equation exceeds the age of
the universe for light ($m_{\nu_h}\approx$10keV) neutrino $\nu
_h$ and relatively large ($\geq$TeV) $\hat{M}$. The life time
can be lowered by lowering the value of $\hat{M}$, but it is
possible to construct models where the decay $\nu
_{h}\rightarrow \nu _{l}+J$ occurs at a much faster rate than
given by eq.(2). Models proposed \cite{v1,v2} in the literature
achieve this by enlarging the Higgs content. They typically
require adding more than one singlet to the minimal seesaw
model.

One of the motivations of the present works is to look
critically at the minimal model with only one singlet Higgs and
three right-handed neutrinos added to the standard model. We
shall show that if one does not insist on the conservation of
total lepton number then it is indeed possible to obtain in this
minimal model a heavy neutrino decay rate much faster than given
in eq.(2). The Majoron in this scenario gets related to a
spontaneously broken horizontal symmetry acting on leptons. It
is possible to choose this symmetry in such a way that the
contribution of the heavy neutrinos to $0\nu \beta\beta$ cancels
naturally. Thus within this minimal model one could accommodate
heavy neutrinos without conflicting with any of the known
constraints.

The second aim of the paper is to see if the heavy neutrino in
this minimal model can be identified with the reported
\cite{simp} 17 keV neutrino. Like any other heavy neutrino, the
17 keV neutrino has to satisfy \cite{lk} constraints imposed by
cosmology (relic densiy and nucleosynthesis), astrophysics (the
supernova SN1987a) and the laboratory experiments
($0\nu\beta\beta$ and oscillations). The desire to satisfy these
constraints  has given rise to ingenious but fairly involved
\cite{models} models for the 17 keV neutrino. We shall show that
the minimal model considered here can meet most of the stringent
constraints. We shall give an explicit example where this
happens. We however find that the mixing angles predicted in
this example do not agree with the laboratory limits on them
coming from neutrino oscillations if the mixing of the 17 keV
neutrino with $\nu _e$ is indeed as large as $\approx 1\%$. One
could avoid this easily by adding an SU(2) triplet of Higgs
field to the minimal seesaw model. The resulting model provides
a much more economical and yet successful description of the 17
keV neutrino than most of the proposed \cite{models} schemes.

In the next section, we discuss the minimal seesaw model and
show that it is possible to obtain a decay rate higher than
given by eq.(2) for a heavy neutrino in this model. The third
section contains a specific example which has fast decay rate
and suppressed neutrinoless double beta decay. The fourth
section summarizes the constraints on the 17 keV neutrino. In
the fifth section we provide a model for the 17 keV neutrino.
The last section summarizes our results.

\section{Heavy Neutrino Decay}
We consider in this section an $SU(2)\times U(1)\times U(1)_H$
model containing three right-handed neutrinos and a complex
$SU(2)\times U(1)$  singlet scalar field $\eta$ in addition to
the standard fields. The right-handed neutrinos are needed to
obtain seesaw masses while $\eta$ as well as a global symmetry
$U(1)_H$ is required to obtain Majoron in the manner suggested
\cite{cmp} by CMP. We shall refer to this model as the minimal
 seesaw model
(MSM). $U(1)_H$ was identified with the total lepton nuber in
reference \cite{cmp}. As was shown \cite{sv} later, this leads
to eq.(2) and hence
 to a slower rate for the decay of heavy neutrino to a lighter
neutrino and Majoron. This decay rate can be increased \cite{v1}
if $U(1)_H$ distinguishes between generations. Explicit models
where this happens \cite{v1,v2} were discussed but this involved
adding more than one singlet scalar field. As we discuss now,
this is unnecessary and one could obtain a fast decay in the
framework of the MSM.

If we do not insist on the conservation of total lepton number
then the most general Yukawa interaction invariant under
$SU(2)\times U(1)\times U(1)_H$ is given as follows:

\begin{equation}
-{\cal L}_Y=\overline{\nu_L'}\frac{m}{<\phi^0>}\;\nu_L'^c\;\;\phi^0
               +\frac{1}{2}\overline{{\nu_R'}^{c}}\left(
                \frac{M_{\eta}}{<\eta>}\eta
                +\frac{M_{\eta^*}}{<\eta^*>}\eta^*
                                        +M_B \right)\nu_R' +H.c.
\end{equation}
$\phi^0 $ is a neutral member of the doublet and
$\nu_L'\;\;,\nu_R'$ are weak eigenstate neutrino fields.
$m$,$M_{\eta}$, $M_{{\eta}^*}$ and $M_B$ are matrices in
generation space. The entries in $m$ are typically of the order
of the charged-lepton or the up-quark masses while those in
$M_{\eta}$,$M_{{\eta}^*}$ and $M_B$ are assumed to be much
larger.  Because of the simultaneous presence of $\eta$ and
$\eta^*$ in eq.(3), ${\cal L}_Y$ cannot conserve total lepton
number. Once lepton number conservation is not insisted upon,
there is no reason to forbid the bare mass term $M_B$ which has
also been included in the above equation. All the terms in
eq.(3) are however required to be invariant under a global
$U(1)_H$ corresponding to some linear combination of the
individual lepton numbers of each generation.

Eq.(3) gives the following mass term for the neutrinos
\begin{equation}
-{\cal L}_{mass}=\frac{1}{2}
 \left( \begin{array}{cc}
\overline{\nu_L'},&\overline{\nu_R'^c} \end{array} \right)
                {\cal M}\left(
                 \begin{array}{c}
                 \nu_L'^c\\
                 \nu_R'
                 \end{array}   \right) +H.c.,
\end{equation}
where
\begin{equation}
{\cal M}=\left(  \begin{array}{cc}
                  0&m\\
                  m^T&M\\
                  \end{array}   \right),
\end{equation}
and
 \begin{equation}
M=M_{\eta}+M_{\eta^*}+M_B.
\end{equation}
for simplicity, we shall assume CP conservation and take all
masses and vacuum expectation values real. Moreover, we shall
work in the seesaw limit, $\hat M \gg  \hat m$, $\hat M,
\hat{m}$ being typical values of the entries in $m$ and $M$
 respectively. Diagonalization of ${\cal M}$ proceeds in the
 well-known \cite{sv} way in
this limit. First, we block diagonalize ${\cal M}$, i.e.,
\begin{equation}
U{\cal M}U^T=\left(
             \begin{array}{cc}
             -mM^{-1}m^T&0\\
              0&M+\frac{1}{2}M^{-1}m^Tm+\frac{1}{2}m^TmM^{-1}\\
                                            \end{array} \right)
              +O\left(\frac{\hat{m}^3}{\hat{M}^2}\right),
\end{equation}
where,
\begin{equation}
U=\left( \begin{array}{cc}
          1-\frac{1}{2}\rho \rho^T&-\rho\\
          \rho ^T&1-\frac{1}{2}\rho^T\rho\\
          \end{array}  \right) +O(\rho^3)
\end{equation}
and $\rho=mM^{-1}$.
$-mM^{-1}m^T$ defines the effective mass matrix $m_{eff}$ for
the light neutrinos in the seesaw limit. Let $O$ be $3\times 3$
matrix which diagonalizes $m_{eff}$
\begin{equation}
Om_{eff}O^T=diag.(\xi_1m_{\nu_1},\xi_2m_{\nu_2},\xi_3m_{\nu_3}).
\end{equation}
$m_{\nu_i}$ ($i$=1,2,3) are positive masses for three light
neutrinos. $\xi_i$ are signature factors which depend upon the
structure of the matrix $m_{eff}$. They can be removed by a
redefinition of the phases of the neutrino fields, i.e.,
\begin{equation}
POm_{eff}O^TP^T=diag.(m_{\nu_1},m_{\nu_2},m_{\nu_3}),
\end{equation}
with
\begin{equation}
P_{ij}=\delta_{ij}e^{\frac{i\pi}{4}(1-\xi_i)}.
\end{equation}
The relation between the weak eigenstates $\nu_L'$,$\nu_R'^c$ and
the light mass eigenstates follows from eq.(10),
\begin{equation}
\nu_{iL}'=[O^TP^*]_{ij}\nu_{jL}+...
\end{equation}
\begin{equation}
\nu_{iR}'^{c}=-[M^{-1}m^TO^TP]_{ij}\nu_{jL}+...
\end{equation}
The terms involving ... contain fields describing the heavy Majorana
neutrinos with masses $\approx O(\hat{M})$. We are primarily
interested in the light fields $\nu_{iL}$.

The vacuum expectation value of $\eta$ breaks the $U(1)_H$ symmetry
spontaneously
giving rise to a Majoron $J$. The latter is related to $\eta$ by
\begin{equation}
J\equiv\;\;\sqrt 2{\rm Im}\eta.
\end{equation}

The couplings of the light neutrinos to the Majoron arise
through couplings of the right-handed fields $\nu_{R}$ to
$\eta$. Using eq.(3),
\begin{equation}
-{\cal L}_J=\frac{i}{2\sqrt 2 <\eta>}\overline{\nu_R '^c}
                        (M_{\eta}-M_{\eta^*})\nu_R'J\,\,\,+H.c.
\end{equation}
In terms of the mass eigenstates $\nu_i$ of the light neutrino
fields, eq.(12,13), we have
\begin{equation}
-{\cal L}_J=\frac{i}{2\sqrt 2 <\eta>}g_{ij}e^{-\frac{i\pi}{4}
             (2-\xi_i-\xi_j)}
                      \overline{\nu_{iL}}{\nu_{jL}}^cJ\;\;\;+H.c.
\end{equation}
where,
\begin{equation}
g_{ij}=[Om_JO^T]_{ij},
\end{equation}
with the Majoron coupling matrix $m_J$ defined as
\begin{equation}
m_J=mM^{-1}(M_{\eta}-M_{\eta^*})M^{-1}m^T.
\end{equation}
In terms of the Majorona fields $\nu_i=\nu_{iL}+{\nu_{iL}}^c$, we
have for any pair $i,j$($i\;\;> j$):
\begin{eqnarray}
-{\cal L}_J&=&\frac{g_{ij}}{\sqrt 2<\eta>}\overline {\nu_i}\nu_j J
\;\;\;\;\;\;\;   \mbox{if}\;\;\;\; \xi_i+\xi_j=0 \nonumber \\
&=&\pm \frac{g_{ij}}{\sqrt 2<\eta>}\overline{\nu_i}\nu_j J
            \;\;\;\;\;\;\;\mbox{if}\;\;\;\; \xi_i+\xi_j=\pm 2.
\end{eqnarray}

The couplings $g_{ij}$ of neutrinos to Majoron are generically
of $O(\frac{\hat{m}^2} {\hat{M}^2})$. However, in specific cases,
the matrix $g$ could be diagonal and the neutrino decay
amplitude may be suppressed. This happens, for example, if $U(1)_H$
 is identified with
the total lepton number \cite{sv}. $M_B$ is zero  and only one
of $M_{\eta}$ and $M_{\eta^*}$ is allowed to be present in
eq.(3). As a result, the Majoron coupling matrix $m_J$
coincides with $m_{eff}$ and $g$ is diagonal. The off-diagonal
Majoron couplings arise \cite{sv} at $O(\frac{\hat{m}^4}
{ \hat{M}^4})$
 in this
case and the decay rate for $\nu_j\rightarrow \nu_i+J$ is
enormously suppressed as in eq.(2)

In general, the off-diagonal couplings of neutrinos to Majoron
arise if any two of the $M_B$, $M_{\eta}$ and $M_{\eta^*}$ are
nonzero.  The matrix $m_J$ is different from $m_{eff}$ in this
case. As long as the $m_J$ does not commute with $m_{eff}$, the
coupling matrix $g$ contains off-diagonal entries at
$O(\frac{\hat{m}^2}{\hat{M}^2})$ leading to a fast decay rate
for $\nu_j\rightarrow\nu_i+J$. In this situation, one could have
heavy neutrinos without any conflict with cosmology. We shall
present a model where this happens in the next section.
\section{A Specific Model}
Following the analysis presented in the earlier section, we now
present an explicit model. We shall make a specific choice of
$U(1)_H$ which not only leads to a fast decay for the heavy
neutrino but also implies vanishing of the neutrinoless double
beta decay amplitude   in a natural manner.  As already
discussed, a fast decay rate can result if $U(1)_H$ allows at
least two of the three possible mass terms
$M_{\eta}$,$M_{\eta^*}$ and $M_B$. This can be done by an
appropriate choice of $U(1)_H$. The requirement of a vanishing
neutrinoless double beta decay amplitude also constrains the
choice of $U(1)_H$. As discussed  by Wolfenstein \cite{w} the
neutrinoless double beta decay amplitude is proportional to the
11 element of the light ($\leq$ O(MeV)) neutrino mass matrix in
the basis which makes the charged lepton mass matix diagonal.
$U(1)_H$ must be chosen to ensure this.

We assume that the ordinary doublet field $\phi$ is neutral
under $U(1)_H$. The Majoron does not have tree-level couplings
to fields other than those of neutrinos in this case. As a
result, the scale of the $U(1)_H$ breaking is not required to be
very high as in some models \cite{as1} with a non-trivial
$\phi$. We shall also require $U(1)_H$ to be vectorial and
assume that no two generations transform identically under
$U(1)_H$. These requirements simplify the task of making the
neutrinoless double beta decay amplitude  vanish.  With these
requirements imposed, the charged lepton mass matrix as well as
the Dirac mass term $m$ in the neutrino mass matrix
automatically become diagonal. The neutrinoless double beta
decay amplitude  then vanishes  if the 11 element of
$m_{eff}\equiv -mM^{-1}m^T$ and hence of $M^{-1}$ is zero. In
the absence of any non-abelian symmetry which relates various
Yukawa couplings, $(M^{-1})_{11}$ can be zero only if $M_{23}$
as well as $M_{22}$ and/or $M_{33}$ vanish. We shall try to be
most general and allow the maximum number of  entries in
$M^{-1}$ to be nonzero. Then, a little consideration shows that
only two choices are possible for $U(1)_H$. These  correspond to
$L_e-3L_{\mu}-L_{\tau}$ and $L_e-3L_{\tau}-L_{\mu}$. We
explicitly discuss the former choice. Analogous considerations
are valid for the latter. With this choice for $U(1)_H$,
$M_{\eta}$, $M_{\eta^*}$ and $M_{B}$ defined in eq.(3) are given
by
\begin{equation}
M_{\eta}=\left(
              \begin{array}{ccc}
              0&M_1&0\\
              M_1&0&0\\
              0&0&M_2\\ \end{array} \right) \end{equation}
\begin{equation}
M_{\eta^*}= \left(
              \begin{array}{ccc}
              M_3&0&0\\
              0&0&0\\
              0&0&0\\ \end{array} \right) \end{equation}
\begin{equation}
M_B=\left(
              \begin{array}{ccc}
              0&0&M_0\\
              0&0&0\\
              M_0&0&0\\ \end{array} \right)\end{equation}
If we parametrize the elements of the diagonal matrix $m$ by
$m_i$ then it follows using eq.(6) that
\begin{equation}
m_{eff}\equiv-mM^{-1}m^T=\left(
              \begin{array}{ccc}
              0&X&0\\
              X&Y&Z\\
              0&Z&W\\ \end{array} \right)
\end{equation}
with
\begin{eqnarray}
X&=&-\frac{m_1m_2}{M_1} \\
Y&=&\frac{m_2^2}{M_1^2M_2}(M_2M_3-M_0^2) \\
Z&=&\frac{m_2m_3M_0}{M_1M_2} \\
W&=&-\frac{m_3^2}{M_2}.
\end{eqnarray}

$m_{eff}$ can be diagonalized by an orthogonal matrix. The zero
entries in $m_{eff}$  imply two independent relations among
three mixing angles and masses. We shall extract these relations
under the physically relevant  (see next section) assumption of
all the mixing angles being small. In this case, $O$ can be
parametrized by
\begin{equation}
O\approx\left(
        \begin{array} {ccc}
        1-\frac{1}{2} (\alpha^2+\beta^2)&\alpha&\beta \\
        -\alpha&1-\frac{1}{2}(\alpha^2+\gamma^2)&\gamma \\
        -\beta&-\gamma&1-\frac{1}{2}(\beta^2+\gamma^2)\\
        \end{array}     \right). \end{equation}
This choice of $O$ as well as eq.(9) lead to the following
relations corresponding to $(m_{eff})_{11}$ and $(m_{eff})_{13}$
being zero respectively:
\begin{equation}
\xi_1m_{\nu_1}(1- \alpha^2 -\beta^2)+\xi_2m_{\nu_2}\alpha^2 +
              \xi_3m_{\nu_3}\beta^2\approx 0
\end{equation}
\begin{equation}
\xi_1m_{\nu_1}\beta-\gamma\alpha\xi_2m_{\nu_2}-\xi_3m_{\nu_3}\beta
 \approx 0.
\end{equation}
Neglecting contribution of $m_{\nu_1}$ in the above equations, we
obtain
\begin{equation}
\frac{\alpha^2}{\beta^2}\approx-\frac{\xi_3m_{\nu_3}}{\xi_2m_{\nu_2}}
\end{equation}
 \begin{equation}
 \frac{\gamma \alpha}{\beta}\approx
 - \frac{\xi_3m_{\nu_3}}{\xi_2m_{\nu2}}.
\end{equation}
Eq.(31) requires $\frac{\xi_3}{\xi_2}$ to be negative. Choosing
$\xi_3=-1$, $\xi_2=\xi_1=1$, the elements of $m_{eff}$  can be
expressed in terms of masses and mixing angles:
\begin{eqnarray}
X&\approx&-\alpha m_{\nu_2}-\beta  \gamma m_{\nu_3}+ \alpha m_{\nu_1}
\nonumber  \\
Y&\approx&\alpha^2 m_{\nu_1}- \gamma^2 m_{\nu_3}+m_{\nu_2}
        (1-\alpha^2-\gamma^2)
\nonumber  \\
Z&\approx&\alpha \beta m_{\nu_1}+ \gamma (m_{\nu_3}+ m_{\nu_2})
\nonumber \\
W&\approx&\gamma^2 m_{\nu_2}+\beta^2 m_{\nu_1}-  m_{\nu_3}
               (1-\beta^2-\gamma^2).
\end{eqnarray}
Eq. (33) allows us to fix the Majoron couplings completely in
terms of mixing angles and masses. With $M_{\eta}$, $M_{\eta^*}$
and $M_{B}$ explicitly given by eqs.(20-22), the couplings
$g_{12}$ and $g_{13}$ responsible for the decay of $\nu_{2,3}$
to $\nu_1$ and a Majoron can be worked out from eq.(17). These
are given by
\begin{eqnarray}
g_{12}&\approx&-2\alpha Y-\alpha\gamma Z -\beta Z \nonumber \\
&\approx&-2\alpha m_{\nu_2}
\nonumber \\
g_{13}&\approx&-\alpha Z+2\alpha \gamma Y+\beta \gamma Z \nonumber
\\
&\approx&\beta m_{\nu_3}
\end{eqnarray}
where we have used eqs.(31-32) and retained only the leading
contributions assuming all the mixing angles to be of similar
magnitudes.  We shall neglect the contribution of $m_{\nu_1}$.
The rates for the decay of $\nu_{2,3}$ are given  by
\begin{eqnarray}
\Gamma(\nu_3\rightarrow\nu_1+J)&=&\frac{(g_{13})^2}{32\pi
           (m_{\nu_3})^3}
                 \frac{(m_{\nu_3}+m_{\nu_1})^2(m_{\nu_3}^2-
                 m_{\nu_1}^2)}
                      {<\eta>^2}  \nonumber \\
           &\approx& \frac{\beta^2}{32\pi}\frac{m_{\nu_3}^3}
            {<\eta>^2},
           \end{eqnarray}
\begin{eqnarray}
\Gamma(\nu_2\rightarrow\nu_1+J)&=&\frac{(g_{12})^2}{32\pi
            (m_{\nu_2})^3}
                 \frac{(m_{\nu_2}+m_{\nu_1})^2(m_{\nu_2}^2-
             m_{\nu_1}^2)}
                      {<\eta>^2}  \nonumber \\
           &\approx& \frac{\alpha^2}{8\pi}\frac{m_{\nu_2}^3}
           {<\eta>^2} .
           \end{eqnarray}
 Analogous result also holds for the decay rate
 of the $\nu_3$ going to $\nu_2$.
The considerations based on the relic density of neutrinos imply
stringent constraints on the lifetime of neutrinos. This is
discussed in \cite{turn}  as a function of the heavy neutrino
mass.  Typically, for a neutrino with mass around 1 MeV, one has
$\tau
\leq 10^9$ sec.
Considerations based on the structure formation in the
universe imply more stringent bound \cite{st} on $\tau$. If the
neutrino lifetime is very long then the
 decay products of the heavy
neutrinos  could  make the universe radiation dominated after
the recombination epoch. If this happens, then the density
perturbations at the time of recombination cannot grow
adequately preventing the  formation of structures. Requiring
that the  life time is short enough for the universe to remain
matter dominated till the present epoch \cite{see} one gests
\begin{equation}
\tau\leq 1.7\times 10^3 sec (\frac{m_\nu}{1\; MeV})^{-2}
\end{equation}
We can use eqs(35,37) to derive the bound on the Majorana mass
scale $<\eta>$
\begin{equation}
<\eta>\;\leq (1.5 \times 10^7 GeV)
(\frac{\beta}{.1})(\frac{m_{\nu_3}}{ MeV})^{1/2}
\end{equation}
Thus a fairly large value is allowed for the scale of the
singlet vacuum expectation value. The situation here is to be
contrasted with the original singlet Majoron model. In this
model the decay rate is typically given by eq.(2) and
the cosmological limit requires
\begin{equation}
<\eta>\;\leq (1.5 \times 10^2 GeV)
(\frac{\beta}{.1})^{1/2}(\frac{m_{\nu_3}}{ MeV})^{3/4}
\end{equation}

Before we close this section,we would like to point out that the
structure similar to eq.(23) for the neutrino mass matrix was
obtained by Valle \cite{v2}. He however did not include both the
terms $\eta$ and $\eta^*$ in the Yukawa couplings but had two
different Higgs scalars to obtain essentially the same
structure.  This extension is not necessary and one could stay
with the minimal structure. His main motivation was to
understand the 17 keV neutrino using the structure(23). This is
not possible as we shall discuss in the next section, unless one
does a delicate fine tuning of the parameters.
\section{17 keV neutrino: Constraints}
The existence of a 17 keV neutrino $\nu_{17}$ mixing
significantly with $\nu_e$ was first reported in 1985.
Subsequent experiments failed to observe it. Last year, there
were further evidences for and against the existence of
$\nu_{17}$ . This has generated a considerable theoretical
interest and various constraints to be satisfied in any model
incorporating $\nu_{17}$ have been worked out in detail
\cite{lk}. The severe constraints mainly come from a) the near
absence of the neutrinoless double beta decay, b) the observed
neutrino signals from the supernova SN1987a, c) the bound on the
relic density of the universe and d) nucleosynthesis. The need
for satisfying all these constraints simultaneously has given
rise to models
\cite{models} for
 $\nu_{17}$
which invoke new physics. As we now discuss, the minimal model
considered above can in fact  meet all the above mentioned
requirements. The detailed prediction on mixing angles made in
the model however disagrees with the known limits coming from
the neutrino oscillation experiments.

One way to incorporate the absence of neutrinoless double beta
decay amplitude is to make $\nu_{17}$ a Dirac particle. The
right-handed component of this Dirac $\nu_{17}$ can either be
sterile or it could be one of the known antineutrinos.  In the
former case, the observed length of the neutrino pulse from the
supernova SN1987a puts  strong constraints \cite{bg} on the
neutrino mass.  This case is marginally allowed, the latest
\cite{tg} limit being around 25 keV.

The alternative in which $\nu_{17}$ is a Dirac particle with a
non-sterile right-handed component can be realized by imposing
\cite{asj2} an unbroken $L_e-L_{\mu}+L_{\tau}$ symmetry.  The
supernova does not imply any restriction in this case. This
scenario is however constrained by cosmological arguments. The
$\nu_{17}$ remains stable in the minimal model with
$L_e-L_{\mu}+L_{\tau}$ symmetry.  The relic density of
$\nu_{17}$ in this model can exceed the cosmological limit
unless annihilation of $\nu_{17}$ into Majorons occurs very
rapidly. But if this annihilation is strong enough, the Majoron
stays in equilibrium till the nucleosynthesis era \cite{lk}.
This again conflicts with the known bound \cite{turn} on the
contribution of the additional spieces to nucleosynthesis.

In the light of the above arguments, we are forced to consider a
majorana  $\nu_{17}$ if we wish to understand it within the
conventional seesaw framework. The MSM has not been seriously
considered as a model for $\nu_{17}$ in the literature because
of difficulty in satisfying the constraint coming from the
neutrinoless double decay and because of the expected \cite{sv}
slow rate for the decay of neutrino. The example presented in
the earlier section shows that this is not the case. This
example in fact satisfies all the above constraints. By virtue
of being a non-sterile Majorana particle, there is no constraint
on $\nu_{17}$ from the supernova. Fast decay avoids conflict
with cosmology. Neutrinoless double beta decay amplitude
nearly vanish in this model and the Majoron does not stay in
thermal equilibrium at the nucleosynthesis era if the scale
$\hat{M}$ exceeds
\cite{f1} about
few GeV. Despite this, the model cannot describe $\nu_{17}$ in a
consistent manner as we now show.

With all three neutrinos nondegenarate and two quite heavy, the
neutrino oscillation experiments require all the three mixing
angles to be small. Adopting the parametrization of Caldwell and
Langacker \cite{lk} , the limits on three mixing angles coming
from the oscillation experiments are given \cite{lk} by
\begin{eqnarray}
|\alpha|\;\;\;<.029& &\mbox{from}\;\;\;\;\nu_e\leftrightarrow
              \nu_{\mu}
\;\;\;\; \mbox{oscillation}\nonumber\\
|\beta|\;\;\;<0.18& &\mbox{from}
\;\;\;\;\nu_e\leftrightarrow\nu_{\tau}\;\;\;\; \mbox{oscillation}
\nonumber\\
|\gamma|\;\;\;<.032& &\mbox{from}\;\;\;\;\nu_{\mu}\leftrightarrow
        \nu_{\tau}
\;\;\;\;\mbox{oscillation} \end{eqnarray}
Only $\beta$ can be as large as reported for the 17 keV
neutrino.  Hence, the $\nu_{17}$ has to be identified with
$\nu_{\tau}$. The absence of the neutrinoless double beta decay
also constraints the masses and mixings. This is already evident
in eq.(31) and implies
\begin{equation}
\alpha^2 \approx \beta^2 \frac{m_{\nu_3}}{m_{\nu_2}}
\end{equation}
With $\beta \approx 0.1$, $m_{\nu_3}\approx 17 keV$ and
$m_{\nu_2}$ corresponding to the experimental limit on
$\nu_{\mu}$ mass, we have
\begin{equation}
|\alpha|\;\;\;>\;\;\;0.025
\end{equation}
Thus one has both the upper and lower limit on $\alpha$. The
model of section 3 contains a nontrivial relation among the
mixing angles. Combining eq.(31) and (32), this is given by
\begin{equation}
|\gamma|\approx |\frac{\alpha}{\beta}|
\end{equation}
The limits expresssed in eq.(40) and (42) are seen to be grossly
incompatible with each other if $\beta$ is around 0.1. As a
result the model fails to describe the observed $\nu_{17}$.

We had used the  $L_e-3L_{\mu}-L_{\tau}$ symmetry in
arriving at the model of section 3. There
exists equivalant model with the symmetry
$L_e-3L_{\tau}-L_{\mu}$. In this model, instead of the 13
element the 12 element of the
effective neutrino mass matrix is zero.
As a  consequence, the relation (43) gets replaced by
\begin{equation}
|\gamma|\approx |\frac{\beta}{\alpha}|
\end{equation}
Unfortunately, this relation is also inconsistent with eq.(40)
in view of the upper bound on $\alpha$. Thus the other choice of
symmetry also does not provide a viable model for $\nu_{17}$.
As already mentioned, these are the only choices of $U(1)_H$
which lead to the absence of neutrinoless double beta decay
amplitude in rather natural manner. Hence, the MSM of the type
described in the earlier section cannot be used to describe
$\nu_{17}$. This conclusion can be evaded if one fine tunes the
parameters \cite{v1}  of the model.

The bounds on mixing angles expressed by eq.(40) are derived
assuming all the neutrinos to be non-degenarate. In case of two
of the neutrinos being almost degenarate, the mixing between
them is not required to be small. In the present case, if
$\nu_{\mu}$ and $\nu_{\tau}$ are degenerate the problematic
bound on $|\gamma|$ no longer holds. The mass matrix in eq.(23)
does not admit this solution unless one fine tunes the
parameters. Specifically, one needs to assume $Y$ and $W$ in
eq.(23) to be  much smaller than the off-diagonal elements.
$m_{eff}$ then displays an approximate $L_e-L_{\mu}+L_{\tau}$
symmetry and one gets a Dirac $\nu_{17}$. Moreover, unlike the
model with a triplet Majoron \cite{asj2}, the heavy neutrino
does have off-diagonal couplings to the Majoron even in the
limit of $W$ and $Y$ becoming  zero. Unfortunately, it is not
possible to choose a $U(1)_H$ which automatically ensures
vanishing of $W$ and $Y$. Therefore these elements  have to be
fine tuned. The required fine tuning is quite delicate
\cite{lk}. In the event of large $\gamma$ the $\nu_{\mu}$
disappearance
\cite{de} experiments  constrain the $|m_{\nu_3}^2-m_{\nu_2}^2|$
to be $\ll$ 0.23 (eV)$^2$.  If $W$ and $Y$ are $\ll$
$X,Z$ then $|m_{\nu_2}-m_{\nu_3}|$ $\approx Y+W$. Thus the
ratio, $|\frac{Y+W}{Z}|$ is required to be $\leq  10^{-9}$. In the
absence of such fine tuning among the parameters, one needs to
enlarge the model of section 3 to accommodate the $\nu_{17}$. This we
do in the next section.
\section{17 keV neutrino: A model}
The basic difficulty in describing $\nu_{17}$  within the scheme
is the relation (43) which is a consequence of $(m_{eff})_{13}$
being zero. We can easily extend the MSM to avoid this. The
extension amounts to  adding an $SU(2)\times U(1)$ triplet
carrying hypercharge $-2$ and transforming trivially under
$U(1)_H$. When the neutral member of the triplet acquires a vev,
the following $6\times 6$ maass matrix results for the neutrinos
\begin{equation}
{\cal M}=\left(
         \begin{array}{cc} \delta m&m\\
               m^T&M\\
                         \end{array} \right),
\end{equation}
where the matrices $m$ and $M$ remain the same as in section
(3). $\delta m$ is now given by
\begin{equation}
\delta m=\left(
         \begin{array} {ccc} 0&0&t\\
                    0&0&0\\
                    t&0&0\\  \end{array}
          \right).
\end{equation}
$t$ refers to the contribution coming from the triplet, which is
assumed to be $\leq O\left(\frac{\hat{m}^2}{\hat{M}}\right)$.
One could block-diagonalize the ${\cal M}$ in the seesaw limit
with the same $U$ as in eq.(8).
\begin{equation}
U{\cal M}U^T=\left(
             \begin{array}{cc}
             m_{eff}&0\\
             0&M\\      \end{array}  \right),
\end{equation}
where
\begin{eqnarray}
m_{eff}&=&\delta m-mM^{-1}m^T \nonumber \\
       &=&\left(
          \begin{array}{ccc}0&X&t  \\
                   X&Y&Z\\
                   t&Z&W\\   \end{array}  \right).
\end{eqnarray}
This $m_{eff}$ differs  from eq.(23) only by the (13) entry. The
former can be diagonalized by an orthogonal matrix $O$ as in
eq.(9) and elements of $m_{eff}$ can be related to mixing angles
and masses. In addition to eq.(33), we now have
\begin{equation}
t=\beta (m_{\nu_1}+ m_{\nu_3})-\alpha\gamma m_{\nu_2}.
\end{equation}
The vanishing of $t$ in the earlier model led to the problematic
relation (43). This is now avoided. The mixing angle and masses
now satisfy only one relation corresponding to the vanishing of
neutrinoless double beta decay amplitude.

The triplet field is neutral under $U(1)_H$. As a result, the
Majoron is still given in terms of ${\rm Im} \eta$. Hence the
Majoron couplings $g_{ij}$ are still given by the basic
expression (eq.(17)) derived in section 2.  The couplings
$g_{12}$ and $g_{13}$ are then given as follows
\begin{eqnarray} g_{12}&\approx &\gamma t-2\alpha Y-\alpha
             \gamma Z-\alpha
              \beta t-\beta Z \nonumber \\
             &\approx &-2\alpha m_{\nu_2}     \end{eqnarray}
\begin{eqnarray} g_{13}&\approx & t(1- \frac{1}{2}(\alpha^2+
               \gamma^2+2\beta^2))
              -\alpha Z+2\alpha \gamma Y
             - \beta^2 Z+\beta\gamma Z \nonumber \\
             &\approx &\beta m_{\nu_3}     \end{eqnarray}
Using the upper limit on $\alpha$ as given in in eq.(40) and
$\beta\approx 0.1$, we get the following decay rates
\begin{eqnarray}
\Gamma(\nu_3\rightarrow\nu_1+J)&=&\frac{(g_{13})^2}{32\pi
            (m_{\nu_3})^3}
                 \frac{(m_{\nu_3}+m_{\nu_1})^2(m_{\nu_3}^2-
                  m_{\nu_1}^2)}
                      {<\eta>^2}  \nonumber \\
           &\approx&(1.5\times 10^{-5} sec^{-1})\left(
            \frac{\beta}{0.1}\right)^2
           \left(\frac{m_{\nu_3}}{10 keV}\right)^3 \left(
           \frac{10^5 GeV}{<\eta>}\right)^2
             \nonumber
           \end{eqnarray}
\begin{eqnarray}
\lefteqn{ \Gamma(\nu_2\rightarrow\nu_1+J)=\frac{(g_{12})^2}{32\pi
          (m_{\nu_2})^3}
                 \frac{(m_{\nu_2}+m_{\nu_1})^2(m_{\nu_2}^2-
          m_{\nu_1}^2)}
                      {<\eta>^2} } \nonumber \\
  &\approx&(5.4\times 10^{-3}sec^{-1})\left(
           \frac{\alpha}{0.03}\right)^2
           \left(\frac{m_{\nu_2}}{100 keV}\right)^3 \left(
         \frac{10^5 GeV}{<\eta>}\right)^2
                    \end{eqnarray}
Analogous result holds for the decay of $\nu_3$ to $\nu_1$ and
the Majoron.  We have chosen $m_{\nu_3}$ to be 10 keV and a
typical value of 100 keV for the muon neutrino mass allowed by
the constraint, eq.(31), coming from the vanishing of
neutrinoless double beta decay amplitude. The scale $<\eta>$ is
arbitrary but as follows from the above equation both the
neutrinos can decay very fast for a large range in this scale.
\section{Summary}
We have considered in this paper, a possibility of describing
heavy neutrinos in a phenomenologically consistent manner within
the MSM.  This is made possible by imposing a global $U(1)_H$
which is chosen in a way that simultaneously ensures the absence
of neutrinoless double beta decay amplitude  and also leads to a
fast decay rate for the heavy neutrinos. It is widely believed
that decay of a heavy neutrino to a lighter neutrino and Majoron
is suppressed in the MSM with only one singlet of Higgs field.
We have shown this not to be the case.

The MSM comes very close to providing the description of the
recently reported 17 keV neutrino. All the astrophysical and
cosmological constraints as well the requirement of the
vanishing neutrinoless double beta decay amplitude are met in
the model. As shown in section (3), the relation among mixing
angles predicted in the model do not however seem to be
satisfied in case of the 17 keV neutrino.  This can be easily
avoided in an extension, which also includes an SU(2) triplet
Higgs field. The major shortcoming of the model is its inability
to solve the solar neutrino problem and to describe 17 keV
neutrino at the same time. The mass difference between the
neutrinos in this case are much larger than required for solving
the solar neutrino problem either through the Mikhyev Smirnov
Wolfenstein \cite{msw} mechanism or through the magnetic moment
\cite{okun} of the neutrino. This would certainly require going
beyond  the conventional seesaw mechanism.


\begin{thebibliography}{99}
\bibitem{f1}If the majoron is to decouple before the temprature
of about 1 MeV, then its coupling to neutrino should satisfy $g
\leq 8\times 10^{-6}$, see ref.\cite{see}.
writing $g\approx \frac{m_{\nu}}{<\eta>}$, it follows that
$<\eta>\;\;\geq 2 GeV$ for $m_{\nu} \sim 17 KeV$.
\end{thebibliography}
\end{document}